\documentclass[conference]{IEEEtran}
\IEEEoverridecommandlockouts
\usepackage{amsmath,amssymb,amsfonts}
\usepackage{algorithmic}
\usepackage{graphicx}
\usepackage{textcomp}
\usepackage{xcolor}
\usepackage{float}
\usepackage{siunitx}
\usepackage[backend=biber, style=ieee, sorting=none]{biblatex}
\usepackage{cases}
\bibliography{myref.bib}
\usepackage[linesnumbered,ruled,vlined]{algorithm2e}
\def\BibTeX{{\rm B\kern-.05em{\sc i\kern-.025em b}\kern-.08em
    T\kern-.1667em\lower.7ex\hbox{E}\kern-.125emX}}

\begin{document}

\title{Highway Discretionary Lane-change Decision and Control Using Model Predictive Control\\
{\footnotesize}
\thanks{This work is supported by Guangzhou basic and applied basic research project under Grant 2023A04J1688.}
}

\author{\IEEEauthorblockN{Zishun Zheng}
\IEEEauthorblockA{\textit{Shien-Ming Wu School of} \\
\textit{Intelligent Engineering} \\
\textit{South China University of Technology}\\
Guangzhou, China \\
202320160023@mail.scut.edu.cn}
\and
\IEEEauthorblockN{Yihan Wang}
\IEEEauthorblockA{\textit{Shien-Ming Wu School of} \\
\textit{Intelligent Engineering} \\
\textit{South China University of Technology}\\
Guangzhou, China \\
wiyihanwang@mail.scut.edu.cn}
\and
\IEEEauthorblockN{Yuan Lin*}
\IEEEauthorblockA{\textit{Shien-Ming Wu School of} \\
\textit{Intelligent Engineering} \\
\textit{South China University of Technology}\\
Guangzhou, China \\
yuanlin@scut.edu.cn}
}

\maketitle

\begin{abstract}
To enable autonomous vehicles to perform discretionary lane change amidst the random traffic flow on highways, this paper introduces a decision-making and control method for vehicle lane change based on Model Predictive Control (MPC). This approach divides the driving control of vehicles on highways into two parts: lane-change decision and lane-change control, both of which are solved using the MPC method. In the lane-change decision module, the minimum driving costs for each lane are computed and compared by solving the MPC problem to make lane-change decisions. In the lane-change control module, a dynamic bicycle model is incorporated, and a multi-objective cost function is designed to obtain the optimal control inputs for the lane-change process. Additionally, A long-short term memory (LSTM) model is used to predict the trajectories of surrounding vehicles for both the MPC decision and control modules. The proposed lane-change decision and control method is simulated and validated in a driving simulator under random highway traffic conditions.
\end{abstract}


\section{Introduction}

The execution of lane-change maneuvers by drivers involves not only the consideration of their own vehicle but also the attention to nearby traffic participants. The MPC method has been proven to be an attractive approach in the domain of autonomous vehicle control \cite{gray2013robust,anderson2010optimal}. MPC uses a dynamic vehicle model to predict future states and determines an optimal control sequence for each moment, aiming to minimize a predefined performance index while satisfying constraints on control and state variables \cite{mayne2000constrained}. Mukai and Kawabe developed a lane-change assistance system based on MPC that formulates the problem of generating optimal lane-change paths as a mixed-integer programming problem, solving it with multiparametric programming \cite{mukai2006model}. With existing infrastructure, environmental measurement and vehicle control can be performed, thereby facilitating lane-change decisions \cite{knoop2017lane}. Utilizing this technology, Kamal et al. designed an Economical Adaptive Cruise Control (EACC) lane-change model capable of real-time selection of the optimal lane for fuel economy \cite{kamal2015efficient,kamal2016efficient}. Tejeddin proposed a Multi-Lane Adaptive Cruise Controller (MLACC), designed to determine the optimal speed and driving lane in real-time using MPC \cite{tajeddin2019ecological}. Compared to Kamal's model, the MLACC model accounts for road grade and lane-change benefit thresholds. However, it only considers the relationship between the ego vehicle and leading vehicles, ignoring the following vehicles. Karlsson outlines a control system for self-driving cars to navigate highway exits, integrating a new trajectory planner and decision manager to enhance safety and comfort \cite{karlsson2019optimal}. Bae introduces a real-time control framework for autonomous lane changes in heavy traffic using cooperative behavior prediction and MPC enhanced by Recurrent Neural Networks in 2020 \cite{bae2020cooperation}. In 2022, he also developed an online framework for smooth-path lane changes in dense traffic, integrating MPC with GANs for maneuver generation, and enhancing performance with adaptive safety measures and noise reduction, validated by simulations for efficiency and safety \cite{bae2022lane}. Ammour addresses autonomous driving safety by presenting an MPC-based collision avoidance algorithm accounting for dynamic traffic, with decision making and trajectory planning simplified for efficiency, and validated through simulations with mixed integer formulations and Sigmoid-based safety constraints \cite{ammour2022mpc}. Bhattacharyya and Vahidi introduces adaptive interactive mixed integer MPC, an adaptive motion planning framework for automated highway merging that optimizes vehicle interactions through mixed integer quadratic programming and inverse optimal control, demonstrating effective merging in simulations \cite{bhattacharyya2023automated}.

When considering the interactions with nearby traffic participants, trajectory of nearby vehicles is an important factor for lane-change decision and control. Trajectory prediction has been proved to be an effective method to improve the accuracy of lane-change decision\cite{jeong2021predictive}. Bae uses a Social Generative Adversarial Network (SGAN) to predict the behavior of multiple drivers in the real-time control framework, which presenting a higher success rate of lane change \cite{bae2020cooperation}. In recent years, significant advancements have been made in the field of trajectory prediction through the application of neural network. Mo introduces a GNN-RNN-based Encoder-Decoder network for interaction-aware trajectory prediction in autonomous driving, which effectively predicts vehicle trajectories in varying traffic scenarios by utilizing RNN for dynamic feature extraction and GNN for inter-vehicle interactions \cite{mo2021graph}. Deo proposes an enhanced LSTM encoder-decoder model with convolutional social pooling for predicting vehicle's motion, offering a multi-modal predictive distribution of future trajectories, showing improvement on accuracy over existing models \cite{deo2018convolutional}.

A controlling method combining lane-change decision and control, inspired by the previous work \cite{tajeddin2019ecological}, is presented here. We propose a two-stage method that sequentially solves the decision-making and vehicle control problems, both of which are addressed by combining MPC and \textit{LSTM with convolutional social pooling} (CS-LSTM) for trajectory prediction. Through this holistic control strategy, an autonomous vehicle can dynamically adapt to the rapidly changing conditions of the road environment, ensuring both the safety and efficiency of lane-change maneuvers.

The paper is organized in the following manner. Section \ref{chapter:Lane-change Decision} describes the lane-change decision method based on MPC. Section \ref{chapter:Lane-change Control} details the establishment of a dynamic bicycle model and the control of the vehicle based on MPC. Section \ref{chapter:Simulation} presents the simulation results and analysis.

\section{Lane-change Decision}
\label{chapter:Lane-change Decision}
When the ego vehicle maintains a straight course on the road and the distance to the lead vehicle in the current lane is less than reference following distance $\Delta s_{ref}$, a lane-change decision can be considered. By comparing the driving cost of the current lane with the cost of the adjacent lanes, the optimal lane for travel is determined, and a decision is made to either change lane or stay in the current lane.
 
\subsection{Modelling}

To address the lane-change decision problem, it is essential to establish a mathematical model for the single-lane car-following problem. In this model, we consider the ego vehicle and the preceding and following vehicles on the current lane as point masses.

For the Adaptive Cruise Control (ACC) problem, take the position of the ego vehicle $s$, the speed of the ego vehicle $v$, the acceleration of the ego vehicle $a$, and the jerk of the ego vehicle $j$ as state variables, and take the acceleration $a$ of the ego vehicle as the control variable. The state space variables and control variables are shown as formulas (\ref{equation:state space variables in ACC})(\ref{equation:control variables in ACC}).

\begin{equation}
\label{equation:state space variables in ACC}
x=\left[\begin{array}{ccccccc}
s & v & a & j
\end{array}\right]^\top
\end{equation}
\begin{equation}
\label{equation:control variables in ACC}
u=a    
\end{equation}

The kinematic equations of ego vehicle are shown in formulas (\ref{equation:s(k+1)})(\ref{equation:v(k+1)})(\ref{equation:a(k+1)})(\ref{equation:j(k+1)}).

\begin{equation}
\label{equation:s(k+1)}
s(k+1)=s(k)+v(k) T_s
\end{equation} 
\begin{equation}
\label{equation:v(k+1)}
v(k+1)=v(k)+a(k) T_s
\end{equation} 
\begin{equation}
\label{equation:a(k+1)}
a(k+1)=u(k)
\end{equation} 
\begin{equation}
\label{equation:j(k+1)}
j(k+1)=\left[a(k+1)-a(k)\right]/ T_s
\end{equation}

\begin{equation}
\label{equation:ACC}
x(k+1)=A x(k)+B u(k),   
\end{equation} 
where

\begin{equation}
A=\left[\begin{array}{cccc}
1 & T_s & 0 & 0\\
0 & 1 & T_s & 0\\
0 & 0 & 0 & 0\\
0 & 0 & -\frac{1}{T_s} & 0
\end{array}\right]
\end{equation}
\begin{equation}
B=\left[\begin{array}{cccc}
0 & 0 & 1 & \frac{1}{T_s}
\end{array}\right] ^\top\\
\end{equation}

Simultaneously, for the sake of convenient narration, several mathematical symbols and their meanings are defined. $\Delta s$ represents the distance between the ego vehicle and the leading vehicle, $\Delta s_f$ denotes the distance between the ego vehicle and the following vehicle, $\Delta s_d$ signifies the desired distance between the ego vehicle and the leading vehicle, and $\Delta s_{d, f}$ indicates the desired distance between the following vehicle and the ego vehicle. At the current moment $k$, the distance between the ego vehicle and the leading vehicle, as well as the following vehicle, can be directly obtained. However, in the predictive time horizon, the distance are acquired through prediction, and the detailed content will be elaborated in subsection \ref{subsection:Future State Prediction}.

To represent the kinematic equations of a vehicle in state-space form, formula (\ref{equation:ACC}) can be obtained. The length of time step $T_s$ is set as 0.1s.

\subsection{Cost Function}

When optimizing for multiple objectives simultaneously, a composite cost function that is the sum of individual terms representing different objectives is commonly used. This approach allows one to balance and prioritize the various aspects of the vehicle's performance that are important for the driving task. The cost function used for lane-change decisions in this paper is presented in formula (\ref{eqation:cost function in MLACC}), and the description of each variable in cost function can be referred to in TABLE \ref{table:lane-change decision}.

\begin{table}[h]
\centering
\caption{Parameters of Cost Function in Lane-change Decision}
\begin{tabular}{ccc}
\hline
\textbf{Parameters} & \textbf{Description}                                                                                                         & \textbf{Value} \\ \hline
$v_{ref}$  & Reference speed                       & 27 m/s         \\
$t_h$               & Time headway during car-following                                                                                            & 1.5 s          \\
$d_0$               & The standstill reference following distance                & 5 m            \\
$\lambda_j$         & Jerk weight term                                                                                                             & 0.1            \\
$\lambda_1$         & Leading vehicle distance weight term                                                                                         & 1              \\
$\lambda_2$         & Following vehicle distance weight term                                                                                       & 0.2            \\
$l_{sw}$     & Lane-change decision                           & 0, 1 or -1         \\
$u_d$               & Desired control effort    &   \\
$J_{th}$              & Lane-change threshold of driving cost & 0.3 \\
$k_p$               & Penalty term on lane change    & 0.1 \\
$t$                 & Current time step                                                                                                            &                \\
$l$                 & Predictive time step                                                                                                         &                \\ 
$T$                 & Length of predictive horizon                                                                                                         & 50               \\ \hline
\end{tabular}
\label{table:lane-change decision}
\end{table}

The initial term of the cost function pertains to the velocity component. Within the scope of the scenario of this study , which necessitates the vehicle's continuous gap-seeking and overtaking maneuvers, the maintenance of a comparatively high traveling speed is of paramount importance. The reference speed $v_{ref}$ is slightly higher than the speed of surrounding vehicles.

\begin{subnumcases}{}
\label{eqation:cost function in MLACC}
&$J= \sum_{l=t}^{t+T}\left|\frac{v\left(l \mid t\right)-v_{ref}}{v_{ref}}\right| +\sum_{l=t}^{t+T} \lambda_j j(l \mid t)$\\
&$\quad+\sum_{l=t}^{t+T} \lambda_1 m_1 +\sum_{l=t}^{t+T} \lambda_2 m_2 $\nonumber\\
\text{s.t.}&$m_1=\left\{\begin{array}{l}
0, \Delta s(l \mid t) \geq \Delta s_d(l \mid t) \\
\frac{\Delta s_d(l \mid t)-\Delta s(l \mid t)}{\Delta s_{ref}}, \\ \Delta s(l \mid t)<\Delta s_d(l \mid t)
\end{array}\right.$ \\
&$m_2=\left\{\begin{array}{l}
0, \Delta s_f(l \mid t) \geq \Delta s_{d, f}(l \mid t) \quad\\ \text {or} \quad l_{sw}=0 \\
\frac{\Delta s_{d, f}(l \mid t)-\Delta s_f(l \mid t)}{\Delta s_{ref}},\\ \Delta s_f(l \mid t)<\Delta s_{d, f}(l \mid t)
\end{array}\right.$ \\
&$\Delta s_d(l \mid t)=d_0+t_h v(l \mid t)$ \\
&$\Delta s_{d, f}(l \mid t)=d_0+t_h v_f(l \mid t)$ \\
&$\Delta s_{ref}=50 \mathrm{~m}$
\end{subnumcases}

The second term represents the jerk of the ego vehicle. However, comfort is not considered the most critical criterion; hence, the weight assigned to this term is relatively low.

$\mathit{m_1}$ represents cost related to the leading vehicle. When ego vehicle is too close to the leader, the driving cost begins to rise. However, when $\Delta s$ is far enough, meaning the actual distance is further than the desired distance $\Delta s_d$, it can be assumed that the leading vehicle does not interfere with the movement of ego vehicle, allowing the vehicle to accelerate freely within its lane. Therefore, the driving cost at this moment is considered to be 0.

When ego vehicle stays in its current lane, the following vehicle should be ignored. Consequently, within the computation of the driving cost for ego vehicle in its lane, the following distance term, denoted as $m_2$, is assigned a value of 0. In contrast, when assessing the driving cost for potential maneuvers into adjacent lanes, the influence of the following vehicle is similar to the $m_1$ scenario.

\subsection{Constraints}

Undoubtedly, in both lane-change and car-following behaviors collision avoidance is vital. Therefore, imposing a minimum longitudinal distance constraint between vehicles is necessary. To define the boundaries of following distance and avoid collision, hard constraints are added:

\begin{subnumcases}{}
\Delta s(l \mid t)>d_{safe}=2b & \\
\Delta s_f(l \mid t)>d_{safe}=2b &
\end{subnumcases}
where $b=5$m represents the body length of the ego vehicle.

Additionally, the constraint for acceleration is:
\begin{equation}
\begin{aligned}
& a \in\left[a_{min}, a_{max}\right]
\end{aligned} \\
\end{equation}
where \( a_{min} = -4.5\, \text{m/s}^2 \), \( a_{max} = 2.6\, \text{m/s}^2 \).


\subsection{Future State Prediction}
\label{subsection:Future State Prediction}

\begin{figure*}[h]
    \centering
    \includegraphics[width=0.65\linewidth]{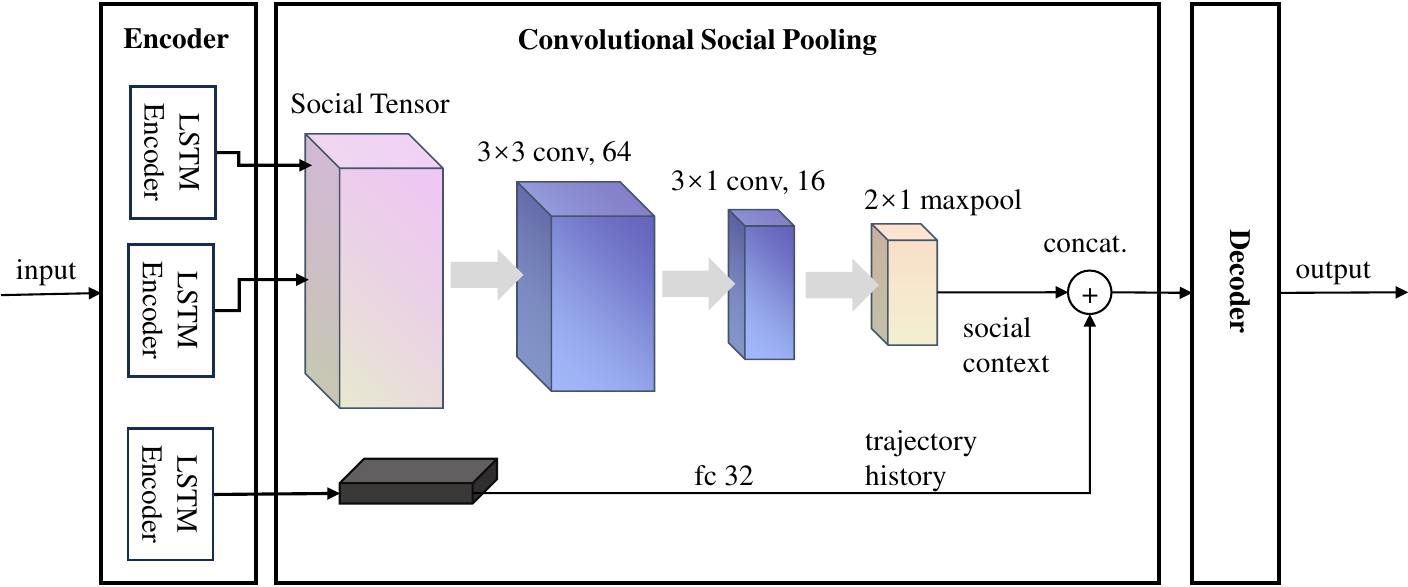}
    \caption{CS-LSTM Model}
    \label{fig:cslstm}
\end{figure*}

To calculate the distance $\Delta s$ between the ego vehicle and the leader in the cost function, future trajectory prediction is necessary. We adopt \textit{LSTM with convolutional social pooling} (CS-LSTM) from \cite{deo2018convolutional} to predict the trajectory of leading vehicle, which is proved to be effective in terms of trajectory prediction.

The input $\mathbf{X}$ of this model is the previous trajectory of predicted vehicle and other vehicles within $110$m over the most recent 30 frames.
\begin{equation}
    \mathbf{X}=\left[\mathbf{x}^{\left(t-30\right)}, \ldots, \mathbf{x}^{(t-1)}, \mathbf{x}^{(t)}\right]    
\end{equation}
where,
\begin{equation}\label{eq10}
    \mathbf{x}^{(t)}=\left[x_{0}^{(t)}, y_{0}^{(t)}, x_{1}^{(t)}, y_{1}^{(t)}, \ldots, x_{n}^{(t)}, y_{n}^{(t)}\right].    
\end{equation}$x_{0}^{(t)}, y_{0}^{(t)}$ are the longitudinal position and lateral position of the predicted vehicle respectively, and other $x,y$ are the longitudinal and lateral position of vehicles $110$m around predicted vehicle.

The output $\mathbf{Y}$ is the predicted trajectory of the leading vehicle in the next 50 frames.
\begin{equation}
    \mathbf{Y}=\left[\mathbf{y}^{(t+1)}, \ldots, \mathbf{y}^{\left(t+50\right)}\right]
\end{equation}
where,
\begin{equation}
    \mathbf{y}^{(t)}=\left[x_{0}^{(t)}, y_{0}^{(t)}\right].
\end{equation}

Fig.\ref{fig:cslstm} illustrates the network architecture of CS-LSTM. It integrates an LSTM encoder, convolutional social pooling layers, and a maneuver-based LSTM decoder. 

The LSTM encoder is utilized to capture the dynamic motion of vehicles. It processes snippets of track history in the most recent 30 frames for each vehicle, and updates LSTM states over the past frames. And LSTMs use shared weights to enable correspondence between LSTM states for all vehicles.

To address the interaction among vehicles, the model then incorporates convolutional social pooling. We establish a $49\times3$ spatial grid to save predicted vehicle and surrounding vehicle locations. In the spatial grid, each column represents a lane and rows are separated by the vehicle length. The social tensor is set up based on the surrounding vehicle information saved in the grid, and is connected to two convolutional layers and a pooling layer. Also, the predicted vehicle’s LSTM state is processed through a fully connected layer.

The above two encodings are then concatenated to form the complete trajectory encoding, which is subsequently passed to a LSTM decoder. The decoder comprises two softmax layers that output probabilities for lateral and longitudinal maneuvers. For each maneuver class, the decoder output future vehicle motion prediction in the next 50 frames. Then we choose the maneuver class with highest probability to calculate the prediction result $Y$.

To demonstrate the effectiveness of using CS-LSTM for prediction, a comparison with another method is necessary. In this paper, the Frozen Time method is employed, which assumes that the surrounding vehicles maintain constant velocity, and their trajectories in the prediction time domain are calculated based on this assumption.

\subsection{Optimal Lane Selection}
At each moment $k$, the ego vehicle retrieves information of surrounding vehicles, including the leading distance $\Delta s_i$, following distance $\Delta s_{f,i}$ and velocity  of the leading vehicles $v_{l,i}$ and the following vehicles $v_{f,i}$ in the left lane $l$, current lane $c$ and right lane $r$, where $i \in \{r,c,l\}$. Subsequently, it first determines whether the left and right lanes exist and whether safety conditions are satisfied for lane change.

Regarding the safety conditions during lane change, considering that the surrounding vehicles maintain a stable speed, a constant value is used as the safety distance criterion, set as $d_{safe}=3b=15$m. Thus, as long as the longitudinal distance with the leading and following vehicles in the adjacent lanes are both greater than $d_{safe}$, it is considered to meet the safety conditions to trigger lane change.

\begin{algorithm}[ht]
\label{Algorithm:lane selection}
\SetAlgoLined
\DontPrintSemicolon
\caption{Determination of $l_{sw}$ and $u_d$}
\KwIn{$\Delta s_i, \Delta s_{f,i}, v, v_{l,i}, v_{f,i}, a, j$}
\KwOut{$l_{sw}$, $u_d$}
\While{\textbf{MLACC engaged}}{
  $(J_c, u_{d,c}) \leftarrow \textbf{NMPC}(\Delta s_c, \Delta s_{f,c}, v, v_{l,c}, v_{f,c}, a, j)$\;
  \uIf{$J_c \leq J_{th}$}{
    \Return $l_{sw} \leftarrow 0, u_d \leftarrow u_{d,c}$\;
  }
  \Else{
    $(J_r, u_{d,r}) \leftarrow \textbf{NMPC}(\Delta s_r, \Delta s_{f,r}, v, v_{l,r}, v_{f,r}, a, j)$\;
    $(J_l, u_{d,l}) \leftarrow \textbf{NMPC}(\Delta s_l, \Delta s_{f,l}, v, v_{l,l}, v_{f,l}, a, j)$\;
    \uIf{$(1+k_p) J_r < J_c$ \textbf{and} $J_r \leq J_l$}{
      \Return $l_{sw} \leftarrow -1, u_d \leftarrow u_{d,r}$\;
    }
    \uElseIf{$(1+k_p) J_l < J_c$ \textbf{and} $J_l < J_r$}{
      \Return $l_{sw} \leftarrow 1, u_d \leftarrow u_{d,l}$\;
    }
    \Else{
      \Return $l_{sw} \leftarrow 0, u_d \leftarrow u_{d,c}$\;
    }
  }
}
\end{algorithm}

For all available lanes in $i$, the solver optimizes the minimum future driving cost in each lane and the corresponding control input using a MPC approach. It then determines the optimal lane at the current moment and executes the decision to either remain in the current lane or change lane, as detailed in Algorithm \ref{Algorithm:lane selection}.

To avoid unnecessary frequent lane changes, a threshold value $J_{th}$ needs to be introduced. Only when the driving cost of the current lane is higher than the threshold value, the lane change decision is considered. Meanwhile, for the driving cost of adjacent lanes, a coefficient $(1+k_p)$ should be multiplied when comparing with the current lane to ensure that the lane change is executed only when there is a significant benefit. Finally, an optimal lane is selected, and the decision value is returned to $l_{sw}$. A value of 0 indicates maintaining the current lane, while 1 indicates a lane change to left and -1 indicates a lane change to right.

\section{Lane-change Control}
\label{chapter:Lane-change Control}
The previous chapter elaborates the methodology for discretionary lane-change decisions at each time step. This chapter will introduce the control mechanisms engaged to maneuver the vehicle during the lane-change process subsequent to the decision, also using the MPC method.

At each discrete time step, the control of the vehicle can be construed as a composite function of acceleration $a$ and the steering angle $\delta$ of the front wheels.

\subsection{Dynamic Bicycle Model}

In a highway scenario, using the dynamic bicycle model can make the calculation of its driving trajectory more accurate. The vehicle dynamics model used in this article is described as formula (\ref{equation:state space variable in control})(\ref{equation:control variable in control})(\ref{equation:state space equation in control}). Six state variables are selected: longitudinal position $x$, lateral position $y$, yaw angle $\varphi$, lateral velocity $u$, longitudinal velocity $v$, and yaw rate $\omega$.
\begin{equation}
\label{equation:state space variable in control}
X=\left[\begin{array}{llllll}
x & y & \varphi & u & v & \omega
\end{array}\right]^\top
\end{equation}
\begin{equation}
\label{equation:control variable in control}
U=\left[\begin{array}{ll}
a & \delta
\end{array}\right]^\top   
\end{equation} %
where 
\begin{equation}
\label{equation:state space equation in control}
\begin{aligned}
 \dot{X}=f(X,U)=\left[\begin{array}{c}
u \cos \varphi-v \sin \varphi \\
v \cos \varphi+u \sin \varphi \\
\omega \\
a+v \omega-\frac{1}{m} F_{Y_1} \sin \delta \\
-u \omega+\frac{1}{m}\left(F_{Y_1} \cos \delta+F_{Y_2}\right) \\
\frac{1}{I_z}\left(l_f F_{Y_1} \cos \delta-l_r F_{Y_2}\right)
\end{array}\right] \\
\end{aligned}
\end{equation}

\begin{figure}[h]
    \centering
    \includegraphics[width=0.75\linewidth]{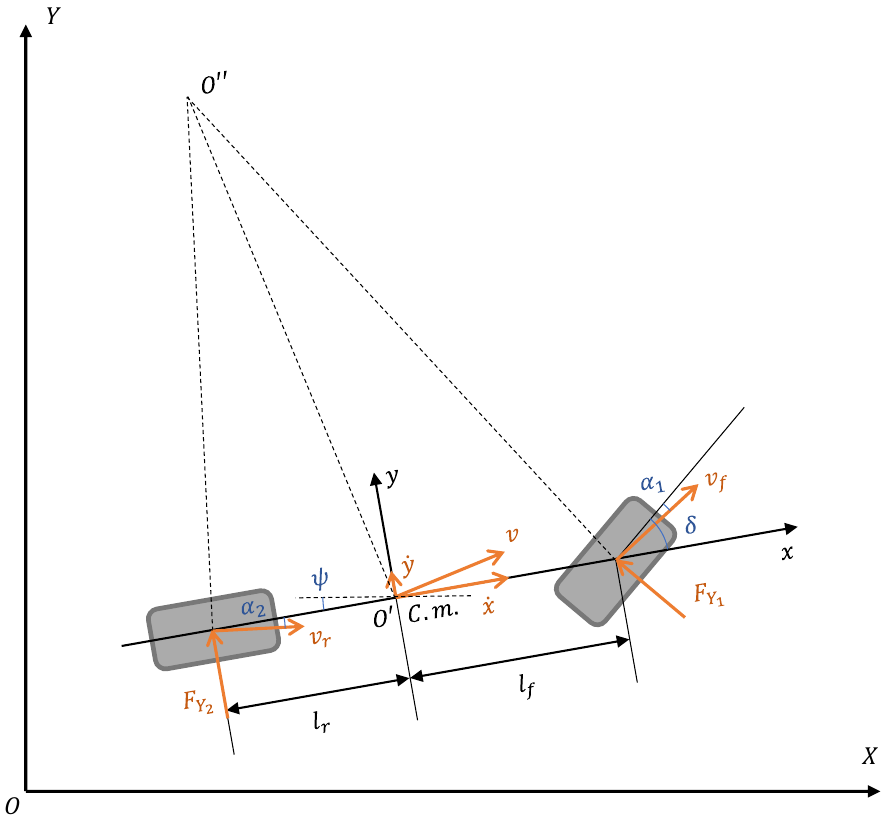}
    \caption{dynamic bicycle model}
    \label{fig:dynamic bicycle model}
\end{figure}

When the front wheel steering angle $\delta$ of the vehicle is very small, it can be approximately obtained:

\begin{align}
& \cos \delta \approx 1 \\
& \sin \delta \approx 0
\end{align}

The lateral forces experienced by the front and rear tires, denoted as $F_{Y_1}$ and $F_{Y_2}$ respectively, can be quantified by considering the tire cornering stiffness $k_f$, $k_r$ and the tire slip angles $\alpha_1$, $\alpha_2$. 

\begin{align}
& F_{Y_1}=k_f \alpha_1 \approx k_f\left(\frac{v+l_f \omega}{u}-\delta\right) \\
& F_{Y_2}=k_r \alpha_2 \approx k_r\left(\frac{v-l_r \omega}{u}\right)
\end{align}

The parameters of the vehicle model are as shown in TABLE \ref{table:lane-change control}.

\begin{table}[h]
\centering
\caption{Basic Parameters of Simulation Vehicle}
\begin{tabular}{ccc}
\hline
\textbf{Parameter} & \textbf{Description}                    & \textbf{Value}         \\ \hline
$m$                & mass of the vehicle                     & 1470 kg                 \\
$k_f$              & front axle equivalent sideslip stiffness & -100000 N/rad           \\
$k_r$              & rear axle equivalent sideslip stiffness  & -100000 N/rad           \\
$l_f$              & distance between $C.m.$ and front axle    & 1.085 m                 \\
$l_r$              & distance between $C.m.$ and rear axle     & 2.503 m                 \\
$I_z$              & yaw inertia of vehicle body             & 2400 kg\(\cdot\)m\(^2\) \\ \hline
\end{tabular}
\label{table:lane-change control}
\end{table}


\subsection{Cost Function}

When the vehicle is driving in its lane without making a lane-change decision, the lateral control objective of the ego vehicle is to maintain the center of the lane, and the longitudinal control objective is to follow the preceding vehicle.

When the vehicle makes a lane-change decision, the reference lateral position $y_{ref}$ is updated to correspond to the centerline coordinates of the targeted adjacent lane. The lateral control aim for the ego vehicle is to converge toward $y_{ref}$, while the longitudinal control objective shifts to maintaining a safe following distance behind the leading vehicle in the target lane.

\begin{equation}
\begin{aligned}
J=&\sum_{l=t}^{t+T} \lambda_v\left\|v(l \mid t)-v_{ref}\right\|^2 \\
& +\sum_{l=t}^{t+T} \lambda_y\left\|y(l \mid t)-y_{ref}\right\|^2 \\
& +\sum_{l=t}^{t+T} \lambda_{\delta}\|\delta(l \mid t)\|^2 +\sum_{l=t}^{t+T} \lambda_a\|a(l \mid t)\|^2\\
& +\sum_{l=t+1}^{t+T-1} \lambda_{\Delta \delta}\|\delta(l \mid t)-\delta(l-1 \mid t)\|^2 \\
& +\sum_{l=t+1}^{t+T-1} \lambda_{\Delta a}\|a(l \mid t)-a(l-1 \mid t)\|^2 \\
& +\sum_{l=t}^{t+T}  \max \left ( 0, \lambda_{l}\left ( {d}_{safe}-{d}_{l}\left (l \mid t\right )\right )\right )\\
\text{s.t.} \quad
& {d}_{l}\left (l \mid t\right ) = \mathbf{y}^{(l)} - s\left (l \mid t\right ) - b
\end{aligned}
\end{equation}

Similar to the cost function in lane-change decision, the ego vehicle needs to maintain a high reference speed $v_{ref}$, to ensure it can overtake other vehicles while driving. Another important factor is the lateral deviation of ego vehicle from the centerline of the lane $y(l \mid t)-y_{ref}$. The designed controller should minimize this deviation to keep the vehicle driving straight. Additionally, for a smooth driving experience, it's necessary to control the acceleration $a$, steering angle $\delta$, rate of change in acceleration(jerk) $\Delta a$, and rate of change in steering angle $\Delta \delta$.

\begin{table}[ht]
\centering
\caption{Weight Parameters of Cost Function}
\begin{tabular}{cccc}
\hline
\textbf{Parameters} & \textbf{Value}           & \textbf{Parameters}      & \textbf{Value} \\ \hline
$\lambda_v$         & 1                        & $\lambda_\delta$         & 100000         \\
$\lambda_y$         & 100                      & $\lambda_{\Delta\delta}$ & 10000          \\
$\lambda_l$         & 500                      & $\lambda_a$              & 1              \\
                    &                          & $\lambda_{\Delta a}$     & 50             \\ \hline
\end{tabular}
\label{table:lane-change control parameters}
\end{table}

Especially, to maintain a safe distance $d_{safe}$ from the leading vehicle, a cost function to regulate the following distance is employed. When the distance to the leading vehicle exceeds $d_{safe}$, the cost remains at 0. If the distance falls below $d_{safe}$, the cost rises sharply in a linear fashion.

\subsection{Constraints}

\begin{itemize}
\item The velocity of the ego vehicle should comply with the speed limit regulations of the highway.
\item The ego vehicle is situated on a three-lane road, and constraints should be applied to the road boundaries.
\item The aforementioned dynamic bicycle model holds true only when the steering angle of the front wheels is small; hence, it is necessary to impose constraints on the values of the front wheel steering angle.
\item Considering the vehicle's actual acceleration and braking capabilities, constraints on acceleration must be applied.
\end{itemize}

\begin{subnumcases}{}
v(l \mid t) \in\left[v_{min}, v_{max}\right] &  \\
y(l \mid t) \in\left[y_{min}, y_{max}\right] & \\
\delta(l \mid t) \in\left[\delta_{min}, \delta_{max}\right] & \quad\\
a(l \mid t) \in\left[a_{min}, a_{max}\right] & 
\end{subnumcases}
where 

$v_{min}=20$ m/s, $v_{max}=30$ m/s;

$y_{min}=-9.6$ m, $y_{max}=0$ m;

$\delta_{min}=-5^\circ$, $\delta_{max}=5^\circ$;

\( a_{min} = -4.5\, \text{m/s}^2 \), \( a_{max} = 2.6\, \text{m/s}^2 \).
\section{Simulation}
\label{chapter:Simulation}
\subsection{Platform \& Rules}
This article uses SUMO(Simulation of Urban MObility) as the simulation platform \cite{krajzewicz2010SUMO}. For surrounding vehicles, the Intelligent Driver Model(IDM) car-following model and LC2013 lane-change model in SUMO are used, canceling the tendency to drive to the right, and overtaking is allowed in all three lanes.

\subsection{Training CS-LSTM}
\begin{figure}[H]
    \centering
    \includegraphics[width=1\linewidth]{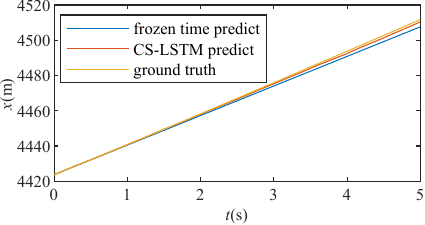}
    \caption{comparison of vehicle longitudinal position prediction}
    \label{fig:flow0001}
\end{figure}
The dataset for training CS-LSTM is generated by traffic flow information from SUMO. The dataset contains $1000$s recorded trajectories of $600$ vehicles in four different highway simulated traffic flows. Fig.\ref{fig:flow0001} shows the comparison of vehicle longitudinal position prediction. When $t=5$s, frozen time prediction has an error of $4.08$m, while CS-LSTM has the error of $1.23$m. This result shows that CS-LSTM has a higher accuracy on trajectory prediction.

\subsection{Result \& Analysis}
\begin{figure}[H]
    \centering
    \includegraphics[width=1\linewidth]{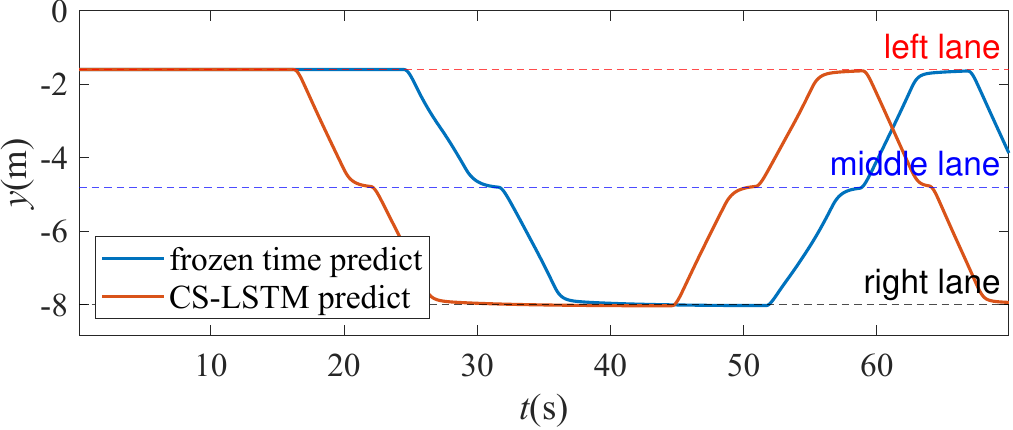}
    \caption{lateral position of ego vehicle}
    \label{fig:lateral_position_compare}
\end{figure}

\begin{figure}[H]
    \centering
    \includegraphics[width=1\linewidth]{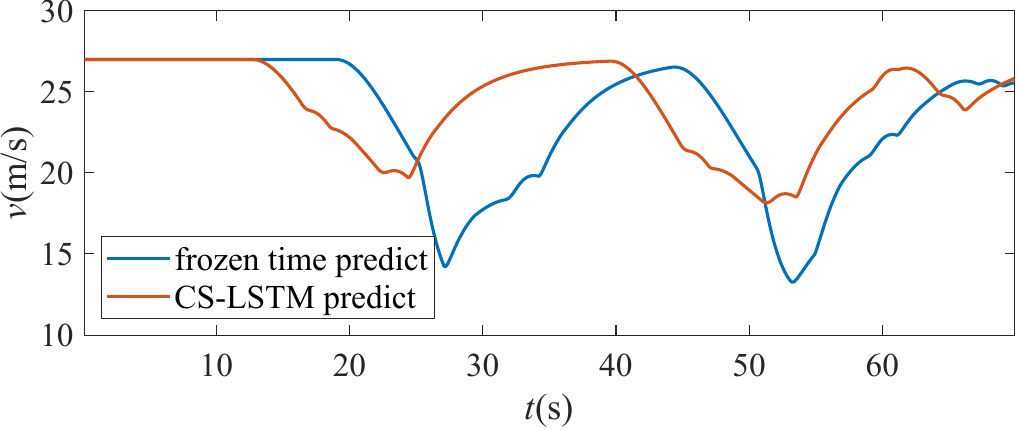}
    \caption{velocity of ego vehicle}
    \label{fig:velocity_compare}
\end{figure}

\begin{figure}[H]
    \centering
    \includegraphics[width=1\linewidth]{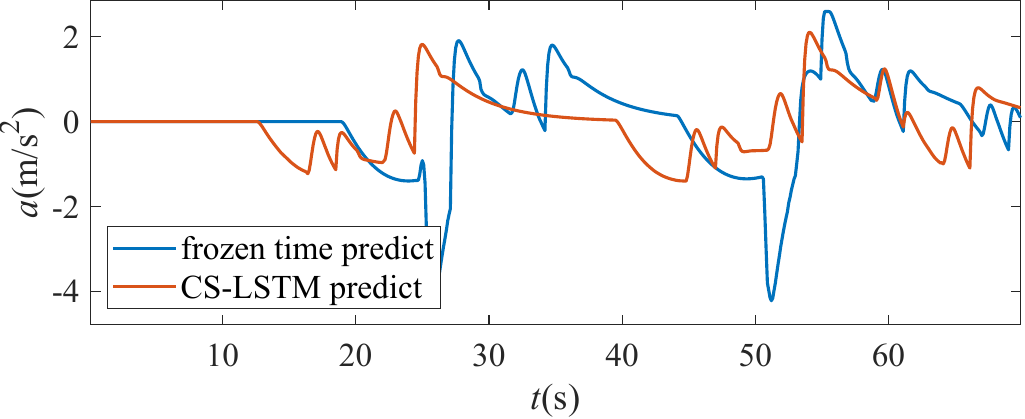}
    \caption{acceleration of ego vehicle}
    \label{fig:acceleration_compare}
\end{figure}

\begin{figure}[H]
    \centering
    \includegraphics[width=1\linewidth]{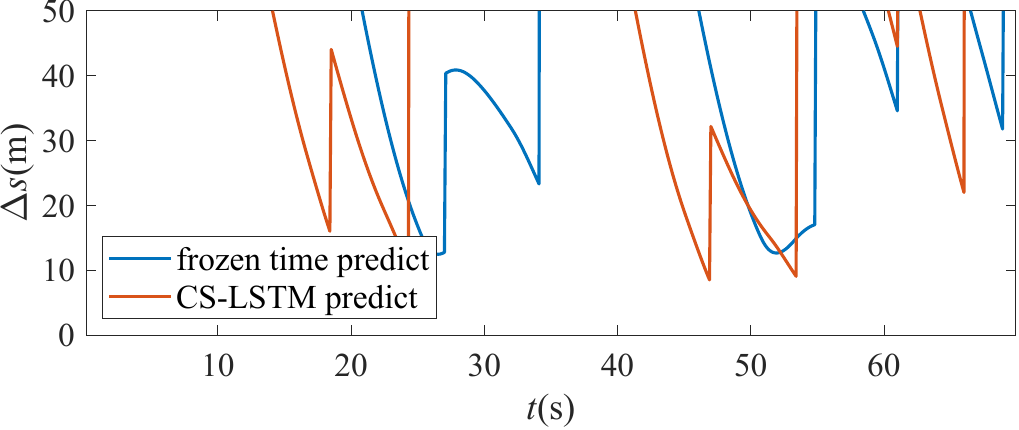}
    \caption{leader distance of ego vehicle}
    \label{fig:leader_distance_compare}
\end{figure}

Figure~\ref{fig:lateral_position_compare}--\ref{fig:leader_distance_compare} illustrates a representative test result, demonstrating the performance of the ego vehicle under the control of the proposed controller. In this road scenario, the ego vehicle demonstrated the ability to identify gaps in the traffic flow and perform intelligent lane changes. When employing both frozen time prediction and CS-LSTM prediction, the ego vehicle encountered highly comparable scenarios and experienced identical decision-making processes. A comparative analysis of the controllers' performance utilizing the two prediction methods in this road test reveals that the controller incorporating CS-LSTM prediction facilitates the ego vehicle in maintaining a improved average velocity. Furthermore, with predictive trajectory of the leading vehicle, it does not necessitate a substantial reduction in speed to guarantee a safe distance when approaching the leading vehicle within the same lane.

\section{Conclusion}
In this study, a control approach is proposed, which combines MPC and CS-LSTM for lane-change decision-making and control of autonomous vehicle on highway. To evaluate the effectiveness of the proposed method, we conducted simulation experiments using SUMO. The results demonstrate that autonomous vehicles equipped with the proposed control method are capable of independently identifying and navigating suitable paths. Moreover, by incorporating CS-LSTM to predict the trajectory of the leading vehicles, the driving performance of the autonomous vehicle is improved. To further validate the effectiveness and robustness of proposed control approach, future research should focus on utilizing simulation software with higher fidelity(such as Carla) and vehicle-in-the-loop test in real-world scenarios.

\printbibliography

\vspace{12pt}

\end{document}